\documentclass{article}

\usepackage[nonatbib,final]{neurips_2024}

\usepackage[utf8]{inputenc} 
\usepackage[T1]{fontenc}    
\usepackage{hyperref}       
\usepackage{url}            
\usepackage{booktabs}       
\usepackage{amsfonts}       
\usepackage{nicefrac}       
\usepackage{microtype}      
\usepackage{xcolor}         

\usepackage[style=numeric-comp, sorting=none]{biblatex} 
\usepackage{amsmath}
\usepackage{amsfonts}
\usepackage{booktabs} 
\usepackage{multirow, makecell}
\usepackage{threeparttable}
\usepackage{xcolor}         
\usepackage{graphicx} 
\usepackage{array}
\usepackage{caption}

\DeclareMathOperator*{\argmin}{arg\,min}
\def\rd{{\textnormal{d}}}
\def\gradlog{{ \nabla \log }}
\def\sbeta{{ \sqrt{\beta_t} }}
\def\dt{{ \mathrm{d} t }}
\newcolumntype{H}{>{\setbox0=\hbox\bgroup}c<{\egroup}@{}}

\newcommand{\sbg}{Schrödinger Bridge}
\newcommand{\ctf}{coarse-to-fine}

\title{A Closer Look at Neural Codec Resynthesis: 
Bridging the Gap between Codec and Waveform Generation}

\author{%
  Alexander H. Liu\textsuperscript{$\dagger$} \quad Qirui Wang\textsuperscript{$\ddagger$} \quad Yuan Gong\textsuperscript{$\dagger$}\thanks{Yuan Gong did this work at MIT, now he is with xAI Corp.} \quad James Glass\textsuperscript{$\dagger$} \\
  \\
  \textsuperscript{$\dagger$} MIT CSAIL \quad \textsuperscript{$\ddagger$} University of Washington \\
  \\
  \texttt{alexhliu@mit.edu}\\
}

\begin{document}
\maketitle

\begin{abstract}
Neural Audio Codecs, initially designed as a compression technique, have gained more attention recently for speech generation. Codec models represent each audio frame as a sequence of tokens, i.e., discrete embeddings. The discrete and low-frequency nature of neural codecs introduced a new way to generate speech with token-based models. As these tokens encode information at various levels of granularity, from coarse to fine, most existing works focus on how to better generate the coarse tokens. In this paper, we focus on an equally important but often overlooked question:  How can we better resynthesize the waveform from coarse tokens? We point out that both the choice of learning target and resynthesis approach have a dramatic impact on the generated audio quality.  Specifically, we study two different strategies based on token prediction and regression, and introduce a new method based on Schrödinger Bridge. We examine how different design choices affect machine and human perception.
Audio demo page: \href{https://alexander-h-liu.github.io/codec-resyn.github.io/}{\texttt{\footnotesize{https://alexander-h-liu.github.io/codec-resyn.github.io/}}}
\end{abstract}

\section{Introduction}

Neural Codec models~\cite{zeghidour2021soundstream,defossez2022high,kumar2024high} initially emerged as compression techniques for audio compression.
Despite being originally proposed for compression, neural audio codec models significantly impacted speech and audio modeling~\cite{wu2024towards} due to their discrete and low-frequency nature.
Having a tokenized representation of audio introduces many benefits.
For example, token-based modeling approaches similar to language models can be adopted for audio generation: VALL-E~\cite{wang2023neural}, Speech-X~\cite{wang2023speechx}, AudioLM~\cite{borsos2023audiolm}, MusicLM~\cite{agostinelli2023musiclm} and MusicGEN~\cite{copet2024simple}, just to name a few.
Besides audio generation tasks, audio codecs can also be applied to cross-modality applications, e.g., making audio and large language model (LLM) integration seamless ~\cite{chen2023lauragpt}.

As shown in Fig.~\ref{fig:overview}, codec models are trained to compress audio into discrete tokens at a low frequency rate to reduce the cost of transmission and storage.
Formally, an encoder would first encode a slice (typically around 10 to 20ms) of signal $s$ into a latent $d$-dimensional embedding $z \in \mathbb{R}^d$.
$z$ will then be iteratively quantized through $N$ Residual Vector Quantization~\cite{juang1982multiple,vasuki2006review} (RVQ) layers into $x_1,x_2,...,x_N$ where
\begin{equation}
    x_i = \operatorname*{\argmin}_{q\in Q_i} \lVert (\underbrace{z - \sum_{j=1}^{i-1} x_j)}_\text{residual}) - q \rVert,
\end{equation}
and $Q_i \in \mathbb{R}^{V \times d}$ is the \textit{codebook} containing $V$ \textit{codes} ($d$-dimensional vectors) of the $i$-th RVQ layer.
Notice how each quantized embedding $x_i$ is a code within the corresponding codebook $Q_i$.
The input audio can therefore be compressed into a sequence of discrete variables which are the indices of the codes in their corresponding codebook.
The bandwidth of neural audio codec models can be controlled by varying the number of codebooks $N$ and the size of each codebook $V$.
The goal of audio codec models is to restore the input signal with the quantized embedding and a decoder such that
$s \approx \text{Decoder}(\sum_{i=1}^{N} x_i).$

Due to the hierarchical structure of RVQ layers, the information carried by the first layer RVQ code $x_1$ is at a coarse level\footnote{In practice, the set of coarse tokens can be defined as the first $c$ RVQ codes $x_{1:c}$. This work considers the most common case $c=1$  without loss of generality since all methods can be extended to $c>1$.}, and that of the remaining layers $x_{2:N}$ gradually becomes more fine-grained~\cite{borsos2023audiolm}.
Recent speech generation models~\cite{wang2023neural,wang2023speechx,borsos2023audiolm,chen2023lauragpt,wang2023viola,rubenstein2023audiopalm} have chosen to prioritize generating the coarse embedding $x_1$.
With the generated coarse embedding $x_1$, the final step to synthesize audio is treated as a separate follow-up question and has received less attention.
Solutions are often ad-hoc, for example, training a coarse-to-fine codec predictor with task-specific information like text and enrolled audio recordings~ ~\cite{wang2023neural,wang2023speechx}, or building a text-and-audio-conditioned codec vocoder~\cite{chen2023lauragpt}.

In this work, we aim to study a question that has been overlooked in codec-based speech generation thus far -- 
How to resynthesize speech using \textit{only} the coarse representation?
We focus on \textit{unconditional} resynthesis that assumes only the coarse embedding $x_1$ is available, and no other task-specific information (such as transcription, speaker, or audio prompt) of the target speech is given.
This assumption allows us to develop general methods not restricted to tasks or data annotation.
We refer to the problem as \textit{Codec Resynthesis} since the ultimate goal is to resynthesize audio from limited codes.
Starting from coarse-to-fine resynthesis, we take a deep dive into \textit{unconditional} codec resynthesis.
With the insight into the learning target, we show how regressing continuous embedding instead of tokens is better.
We further improve the modeling approach, introducing a discrete-to-continuous Codec \sbg.
Finally, we present the strengths as well as the limitations of different methods, along with more challenges in codec resynthesis.

\begin{figure*}[!t]
    \vspace*{0pt} 
    \begin{minipage}[t]{0.55\textwidth}
        \input{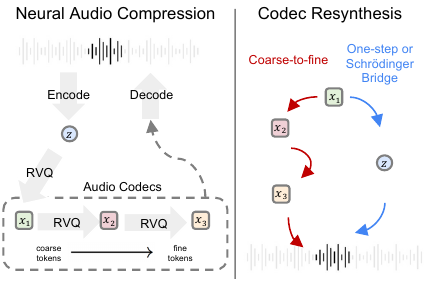}
    \end{minipage}
    \hspace{0.02\textwidth}
    \begin{minipage}[t]{0.45\textwidth}
        \vspace*{0.08\textheight}
        \input{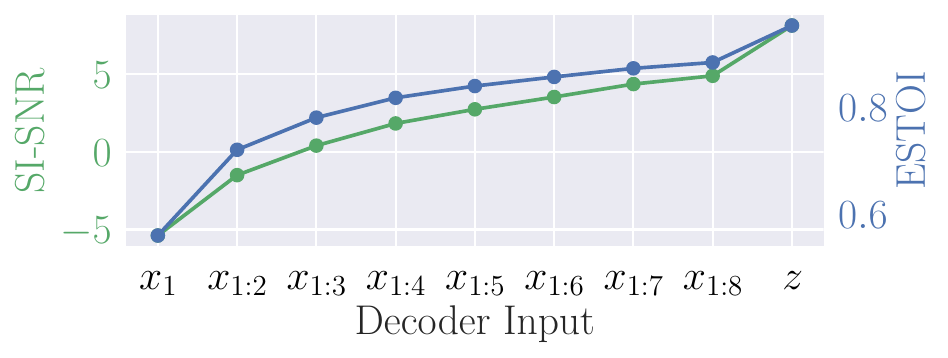}
    \end{minipage}
    \vspace{-20pt}
\end{figure*}

\vspace{-5pt}
\section{Neural Codec Resynthesis}
\label{sec:method}
\vspace{-5pt}

We consider codec resynthesis at the sequence level with non-autoregressive models, i.e., generating complete speech from the sequence of discrete embeddings $(x_1^{(1)},...,x_1^{(l)},...,x_1^{(L)})$ where $L$ is the sequence length.
We shorthand $x_{i} \equiv x_i^{(1:L)}$ hereafter for simplicity.

\vspace{5pt}
\noindent{\textbf{Coarse-to-fine Resynthesis.}}
Due to the hierarchical structure of RVQ layers, each RVQ code $x_i$ depends on all the codes from prior layers $x_{1:i-1}$.
A simple method for codec resynthesis is therefore to \textit{iteratively} predict the RVQ codes $x_{2:N}$ given the first $x_1$.
This can be achieved by training a model, parameterized by $\theta$, to maximize 
\begin{equation}
    \sum_{i=2}^{N} p_\theta(x_i|x_{1:i-1},i)
\end{equation}
where $p_\theta(x_i|\cdot)$ is a categorical distribution over codebook $Q_i$.
The process can be viewed as a coarse-to-fine prediction since the later RVQ codes encode fine-grained audio information.
We note that similar coarse-to-fine models have been studied in prior works where they are autoregressive~\cite{borsos2023audiolm} or text-and-audio-conditioned~\cite{wang2023neural,wang2023speechx}.

\vspace{5pt}
\noindent{\textbf{Is predicting the remaining RVQ codes $x_{2:N}$ necessary?}}
While it is reasonable to resynthesize speech in a coarse-to-fine manner, modeling codes from all RVQ layers introduces a multi-task learning overhead during training and the risk of error propagation during inference.
Prior work~\cite{copet2024simple} attempted to address the issue through parallel prediction but found coarse-to-fine prediction worked best empirically.

To find an alternative, we take a closer look at the quality of the audio decoded from representations at different layers of the codec model.
Results are shown in Fig.~\ref{fig:sisnr}.
As expected, audio quality gradually improved when involving more fine-grain embeddings.
However, a key observation is that the pre-quantized embedding $z$, although never used as decoder input during training, yields the best audio quality.
Since the ultimate goal is to generate high-fidelity audio, this observation suggests that predicting the remaining RVQ codes $x_{2:N}$ may \textit{not} be necessary.

\vspace{5pt}
\noindent{\textbf{One-step Resynthesis.}}
In light of our previous finding, we can train a one-step resynthesis model that simply predicts $z$ directly from $x_1$ through a regression model $f_\theta$ that minimizes 
\begin{equation}
    \lVert f_\theta(x_1) - z  \rVert.
\end{equation}
We refer to this regression-based method as one-step resynthesis since projecting $x_1$ to $z$ requires only a single forward pass of the model, and the result can be directly applied for decoding.
Conversely, the coarse-to-fine method requires $N-1$ iterations to acquire $x_{2:N}$.
Prior work has found one-step resynthesis to be beneficial with audio and text conditions available~\cite{chen2023lauragpt}, but it is unclear whether unconditional resynthesis is possible.

\vspace{5pt}
\noindent{\textbf{Codec \sbg~Resynthesis.}}
Although one-step generation sounds appealing,
recent trends in generative models suggested iterative models, such as diffusion model~\cite{ho2020denoising}, tend to synthesize data of better quality.
From the output audio of one-step resynthesis (see demo page), we also find it results in robotic-sounding artifacts in the speech.
This motivated us to explore iterative methods operating in the continuous embedding space to learn the mapping between $x_1$ and $z$.

The \sbg~(SB) problem~\cite{schrodinger1932theorie} aimed to find the entropy-regularized optimal transport between two arbitrary distributions, $p_{x_0}$ and $p_{x_1}$.
The solution to SB can be expressed by the following forward~(\ref{eq:fsb}) and backward~(\ref{eq:rsb})~stochastic differential equations (SDEs)~\cite{chen2021likelihood}:
\begin{subequations}
    \begin{align}
        \rd x_t &= [f_t + \beta_t~\gradlog {\Psi}_t ] \dt + \sbeta\rd W_t, x_0 \sim p_{x_0}, \label{eq:fsb}
        \\
        \rd x_t &= [f_t - \beta_t~\gradlog \hat{{\Psi}}_t ] \dt + \sbeta\rd \overline{W}_t, x_1 \sim p_{x_1}, \label{eq:rsb}
    \end{align}
    \label{eq:sb-sde}
\end{subequations}
where the $T$-step stochastic process is represented by $x_t$ with $t \in \{0,\frac{1}{T},\frac{2}{T},...,1\}$, $f_t(x_t)$ is the linear drift, $\beta_t \in \mathbb{R}^d$ is the diffusion, $W_t,\overline{W}_t \in \mathbb{R}^d$ are the standard and reversed Wiener process. 
The terms $\gradlog {\Psi}_t(x_t)$ and $\gradlog \hat{{\Psi}}_t(x_t)$ are the forward and backward non-linear drifts for SB with $\Psi,\hat{{\Psi}}$ being the solution to the following coupled PDEs
\begin{equation}\label{eq:sb-pde}
\begin{cases}
   \frac{\partial \Psi}{\partial  t} = - \nabla \Psi^\top f - \frac{1}{2} \beta \Delta \Psi \\
   \frac{\partial \hat{{\Psi}}}{\partial  t}  = - \nabla \cdot (\hat{{\Psi}} f) + \frac{1}{2} \beta \Delta \hat{{\Psi}}
\end{cases} s.t. \Psi_0\hat{{\Psi}}_0 = p_0, \Psi_1 \hat{{\Psi}} = p_1.
\end{equation}
SB provides a general mathematical framework for distribution mapping, but solving it can be challenging in practice since $\Psi$ and $\hat{{\Psi}}$ are often intractable in real world applications.

Fortunately, prior work has shown that SB can be tractable for certain applications where paired data of the two distributions is available~\cite{liu2023i2sb}.
By setting $f:=0$ (merging linear drift into non-linear drift) and $\hat{{\Psi}}_0 := \delta_{x_0}$ (Dirac delta distribution centered around $x_0$), 
a neural network $\epsilon_\theta$ for estimating the score function $\gradlog \hat{{\Psi}}_t$ of the backward SDE (\ref{eq:rsb}) can be trained through minimizing
\begin{equation}
\label{eq:sb-loss}
    \lVert \epsilon_\theta(x_t,t) - \frac{x_t-x_0}{\sigma_t}  \rVert,~~
    x_t \sim \mathcal{N}(~\frac{\bar{\sigma}_t^2}{\bar{\sigma}_t^2+{\sigma}_t^2}x_0 + \frac{{\sigma}_t^2}{\bar{\sigma}_t^2+{\sigma}_t^2} x_1~,~ \frac{{\sigma}_t^2\bar{\sigma}_t^2}{\bar{\sigma}_t^2+{\sigma}_t^2}\cdot I~),
\end{equation}
with $\sigma^2_t {:=} \int_0^t \beta_\tau \rd \tau$ and $\bar{\sigma}^2_t {:=} \int_t^1 \beta_\tau \rd \tau$.

\begin{table*}[!t]
    \centering
    \caption{
    {\fontsize{8pt}{10pt}\selectfont 
        Codec resynthesis results.
        All metrics are the higher the better except WER, see \S\ref{subsec:setup} for detailed explanation.
        Best results are \textbf{bolded}.
        Number of function evaluations (NFE) reflects the number of forward passes required for synthesis.} 
    }
    \label{tab:ll}
    \vspace{-5pt}
    \resizebox{0.87\textwidth}{!}{  
    \begin{threeparttable}
        \begin{tabular}{lccccHccc}
        \toprule
        \multirow{2}{*}{Decoder input} & \multirow{2}{*}{NFE} & \multicolumn{3}{c}{Intrusive Metrics}  & & \multicolumn{3}{c}{Perceptual Metrics} \\
        \cmidrule{3-5}
        \cmidrule{7-9}
        & & SI-SNR ($\uparrow$) & ESTOI ($\uparrow$) & ViSQOL ($\uparrow$) & & WER(\%; $\downarrow$) & SIM ($\uparrow$) & MOS ($\uparrow$) \\
        \midrule
        \midrule
        \multicolumn{9}{l}{Baseline} \\
        ~ 1$^\text{st}$ RVQ code $x_1$ &   & -5.39 & 0.56 & 2.99 & & 24.7 & 0.217 & - \\
        \midrule
        \multicolumn{8}{l}{Resynthesis Methods} \\
        ~ Coarse-to-fine  & 7 & -3.09 & 0.71 & 3.36 & & 22.2 & 0.469 & 3.19 {\footnotesize $\pm$ 0.12} \\
        ~ One-step regression & 1 & \textbf{-1.12} & \textbf{0.75} & \textbf{3.52} & & \textbf{11.5} & 0.495 & 3.06 {\footnotesize $\pm$ 0.13}\\
        \cmidrule{2-8}
        ~ \multirow{5}{*}{\sbg} & 1 & -1.38 & 0.74 &3.49 & & 14.5 & 0.491 & 2.95 {\footnotesize $\pm$ 0.14} \\
        & 4 & -1.55 & 0.74 &3.47 & & 18.3 & \textbf{0.507} & - \\
        & 7 & -1.90 & 0.73 &3.42 & & 19.7 & 0.506 & 3.43 {\footnotesize $\pm$ 0.11} \\
        & 16 & -2.30 & 0.72 &3.37 & & 21.1 & 0.501 & \textbf{3.46 {\footnotesize $\pm$ 0.11}} \\
        & 32 & -2.52 & 0.71 &3.33 & & 22.2 & 0.493 & - \\
        \midrule
        \multicolumn{8}{l}{Topline} \\
        ~ 8 RVQ code $x_{1:8}$ &   & 4.34 & 0.88 & 4.27 & & 2.7 & 0.861 & - \\
        ~ Pre-quantize emb. $z$ &   & 4.87 & 0.95 & 4.55 & & 2.7 & 0.922 & - \\
        ~ Ground Truth &   & - & 1.00 & 5.00 & & 2.4 & 1.000 & 3.74 {\footnotesize $\pm$ 0.11} \\
        \midrule
        \bottomrule
        
    \end{tabular}
    \end{threeparttable}
    }
    \vspace{-10pt}
\end{table*}

\vspace{3pt}
For codec resynthesis, we are interested in transporting between the distribution $p_{x_0} \equiv p_z$  of pre-quantized embedding $x_0 \equiv z$  and the first RVQ code distribution $p_{x_1}$.
Since the pair relation $(x_0,x_1)$ is available through the audio encoding process, Codec \sbg~can be trained directly with Eq.~\ref{eq:sb-loss}.
During inference, Codec \sbg~can be used to construct the backward SDE (Eq.~\ref{eq:rsb}) and derive pre-quantized embedding $x_0$ from $x_1$ using DDPM~\cite{ho2020denoising}.
The backward process can be simulated with different step sizes by breaking down the schedule from $t=1$ to $t=0$ into more/less segments, traversing iteratively from $x_1$ to $x_0$.
In practice, smaller step sizes result in better generation quality~\cite{de2021diffusion,chen2021likelihood,liu2023i2sb} at a cost of more forward passes through the model.

\vspace{-6pt}
\section{Experiments}
\vspace{-4pt}

Due to space constraints, details on model architecture, training hyperparameters, datasets, and evaluation metrics are provided in Appendix Section~\ref{subsec:setup}.
In Table~\ref{tab:ll}, we report resynthesis results on LibriSpeech~\cite{panayotov2015librispeech} \texttt{\footnotesize test-clean}.
The performance baseline is obtained from audio decoded from the first RVQ code $x_1$ without resynthesis.
For resynthesis methods, the decoder of Encodec takes the resynthesis model output as input.
Different toplines should be considered for different methods:
(1) decoding with full RVQ codes $x_{1:8}$, which is the topline for \ctf~method;
(2) decoding with pre-quantized embedding $z$, which is the topline for one-step regression and \sbg.
Ground truth is the raw audio used as the reference for intrusive metrics.

\noindent\textbf{The pre-quantized embedding is consistently a better target.}
As in the findings shown in Fig~\ref{fig:sisnr}, the results in Table~\ref{tab:ll} indicate that pre-quantized embedding $z$ not only provides a higher performance upper bound (see toplines), but also results in better models when used as a learning target.
The one-step regression model performs better on all intrusive metrics at a significantly lower inference cost (NFE=1).
In addition, \sbg~consistently performs better than the \ctf~model in both objective and subjective metrics at the same (NFE=7) or lower (NFE=4) inference cost.
In short, pre-quantized representations of codec models are better than tokens for resynthesis.

\noindent\textbf{A good objective score does \textit{not} imply better audio quality.}
It is worth noting that the one-step regression model is considered the best model in terms of WER and all intrusive metrics in Table~\ref{tab:ll}, including SI-SNR and ViSQOL that are commonly adopted for codec model development~\cite{defossez2022high,kumar2024high}.
However, this result contradicts human perception as reflected by MOS.
In fact, resynthesized speech from one-step regression exhibited a lot of artificial sounding and robotic voices (see audio samples on the demo page), resulting a similar MOS to the \ctf~model.
In contrast, Codec \sbg~significantly reduced the artifacts when taking a smaller step size (more NFE as a trade-off), resulting in more natural output as reflected by MOS.
This finding suggests that the well-known concept that denoising methods like diffusion~\cite{ho2020denoising} are strong in generating high-fidelity data also holds for codec resynthesis.

\begin{figure*}[!t]
\begin{minipage}[t]{0.48\textwidth}
    \input{figure/nfe}  
\end{minipage}
\hfill
\begin{minipage}[t]{0.48\textwidth}
    \vspace*{0.05\textheight}  
    \centering
\vspace{-32pt}
\captionof{table}{{\fontsize{8pt}{10pt}\selectfont Codec resynthesis result with different amounts of training data. LibriLight~\cite{librilight} (LL) contains more than 60k hours of speech. It is more than 62.5x larger than LibriSpeech~\cite{panayotov2015librispeech} (LS) which contains 960 hours of speech only.}}
\label{tab:ls}
\begin{threeparttable}
    \resizebox{0.8\textwidth}{!}{  
        \begin{tabular}{lccHcc}
        \toprule
        \multirow{3}{*}{Decoder input} & \multicolumn{5}{c}{Training data} \\
        \cmidrule{2-6}
         & \multicolumn{2}{c}{LL (60k hours)} & & \multicolumn{2}{c}{LS (960 hours)} \\
        \cmidrule{2-3} \cmidrule{5-6}
        & WER(\%) & SIM & & WER(\%) & SIM \\
        \midrule
        Coarse-to-fine  & 22.2 & 0.469 & & 24.6 & 0.435 \\
        One-step        & \textbf{11.5} & 0.495 & & 28.8 & 0.233 \\
        SB {\footnotesize (NFE=1)} & 14.5 & 0.491 & & \textbf{16.0} & 0.482 \\
        SB {\footnotesize (NFE=7)} & 19.7 & \textbf{0.506} & & 22.7 & \textbf{0.485} \\
        \bottomrule
        \end{tabular}
    }
\end{threeparttable}
\vspace{-5pt}  
\end{minipage}
\vspace{-5pt}
\end{figure*}

\noindent\textbf{Iterative methods improve sound quality, not content.}
Next, we are interested in finding out the reason why iterative methods provided a better MOS yet worse WER.
Interestingly, we found that speech intelligibility (as measured by Whisper) does \textit{not} increase as a function of NFE as shown in Fig.~\ref{fig:nfe}.
This is surprising since single NFE degenerates these iterative methods into one-step methods.
For the \ctf~model, single NFE predicts only the second RVQ code $x_2$, which suggests that the (predicted) fine-grain representation $x_{3:8}$ actually introduces more noise than signal with regard to content.
Single NFE \sbg~is conceptually the same as the one-step regression model, which results in 14.5\% WER that is closer to that of the latter 11.5\% (Table~\ref{tab:ll}).
This indicates that the content of speech is harder to preserve through a more sophisticated backward process.
We note that the problem could potentially be solved with ad-hoc model design for specific applications, e.g., conditioning the \ctf~model with phone sequence~\cite{wang2023neural}.
On the other hand, speaker similarity generally improved as NFE increased, and we hypothesize that naturalness plays an important role.
The difference in reference-free MOS between different NFEs with \sbg~in Table~\ref{tab:ll} also supported this hypothesis. 

\noindent\textbf{Iterative methods are less prone to overfitting.}
In Table~\ref{tab:ls}, we report the results of training models on LibriSpeech (960 hour) instead of LibriLight (60k hour) to observe model behavior when massive training data is not available.
Surprisingly, the one-step method suffered significantly from the lack of training data, resulting in the worst WER and SIM.
In contrast, the coarse-to-fine and \sbg~models have significantly less loss in performance.
In practice, we also found that the one-step regression model overfit the smaller training set after 300k updates (best result reported), whereas both the coarse-to-fine and \sbg~model consistently improved throughout the 500k training steps.
These results suggest that iterative methods that partition the task into smaller subtasks are less prone to overfitting and more robust to low-resource scenarios.

\vspace{-5pt}
\section{Conclusion}
\vspace{-4pt}

\noindent\textbf{Summarizing different methods.}
We first conclude that among the methods explored in this work, \ctf~generation is the least ideal method with poor audio quality and high inference cost.
The one-step regression model is efficient and effective in subjective metrics but results in worse objective scores and robustness.
Codec \sbg~falls slightly behind but can be particularly useful for generating high-fidelity audio.
Finally, we note that while a clear improvement is made by all three methods over the baseline, results suggest that these methods still have much room for improvement.

\noindent\textbf{Rethinking the evaluation metrics.}
We note that the ultimate goal of codec resynthesis is finding the mapping between a coarse-level codec and realistic audio, which can be a one-to-many mapping.
In other words, any resynthesized codec embedding that decodes into realistic audio should be considered a success, the ``ground truth'' should only be used for training but not evaluation.
This is especially true for applications like text-to-speech.
Therefore, intrusive metrics are \textit{not} ideal for the task. 
While reference-free machine perceptual metrics like WER can be a good alternative, it does not reflect human preference.
Finding a better automatic subjective metric that better reflects human preference is still an important open question.

\noindent\textbf{Revisiting token-based and bridge-style generation.}
Recent token-based models~\cite{wang2023neural,wang2023speechx} and bridge-style models~\cite{le2024voicebox,liu2023generative,vyas2023audiobox} each have their strengths in speech generation.
While they appear as two distinct concepts, our findings suggest it is possible to combine the advantages of the two different methods, as demonstrated by the proposed Codec \sbg.

\noindent\textbf{Limitations.}
We left exploring the generalizability of codec resynthesis as an important future work.
For example, the multilingual setup is expected to be even more challenging.
Besides speech, sound and music are also common uses of audio codec models.
Generalizing codec resynthesis beyond speech can potentially reduce the burden of more audio generative models, e.g., music generation model~\cite{copet2024simple} that have to autoregressively generate codec both left-to-right and course-to-fine.

\noindent\textbf{Acknowledgments.}
{\footnotesize The authors would like to thank Guan-Horng Liu for the helpful discussion, Roham Mehrabi and Pei-Ling Chiang for setting up human evaluation on AmazonTurk.}

\newpage
\printbibliography

\newpage
\appendix
\section{Appendix}

\subsection{Experiment Setup}
\label{subsec:setup}
\vspace{3pt}
\noindent\textbf{Model.}
The 6kbps Encodec~\cite{defossez2022high} is used for the codec model with $d=128$ dimensional embedding at 75Hz, $N=8$ RVQ layers each with a codebook of size $V=1024$.
For all three methods, the input is the discrete embedding $x_1$ of the first Encodec RVQ layer.
We trained a 12-layer Transformer Encoder~\cite{vaswani2017attention} with a 16-head self-attention, an embedding dimension of 1024/4096 for self-attention/feedforward layers, and a 0.05 layer dropout.
For the \ctf~model, we used a learnable stage embedding to encode the state $i$.
For the \sbg~model, we used sinusoidal positional embedding~\cite{vaswani2017attention} to encode the timestep $t$.
Adaptive LayerNorm~\cite{xu2019understanding} is used to normalize the output of each layer conditioning on the stage or time embedding.
We used the Adam optimizer with a weight decay of 0.01 to train each model for 1M steps, with learning ramping up in 32k steps and then linearly decaying to 0 for the remaining steps.
The peak learning rate for each method is swept, 1e-4/5e-4/5e-4 is used for the \ctf/one-step/SB~model based on validation loss.
Models have around 210M parameters, and training takes 9 days on 4 A6000 GPUs.
For the \sbg~model, we set $T=1000$ and $\beta_t$ followed a symmetric noise scheduling~\cite{de2021diffusion} peaking at 0.3.
Additionally, we found that conditioning the model on initial point $x_1$ (by projecting and adding it to the Transformer input) regardless of $t$ helpful.

\vspace{3pt}
\noindent\textbf{Data.}
All models are trained on LibriLight~\cite{librilight}, an audiobook corpus with 60k hours of English speech.
The \texttt{\footnotesize dev-clean} / \texttt{\footnotesize test-clean} subset of LibriSpeech~\cite{panayotov2015librispeech} is used for validation/evaluation respectively.
Each audio is randomly cropped to equal length to form an 800-second batch.
 
\vspace{3pt}
\noindent\textbf{Evaluation Metrics.}
To assess the quality of the resynthesized speech, the following metrics were used: Scale-Invariant Signal-to-Noise Ratio (SI-SNR);
Extended Short-Time Objective Intelligibility (ESTOI;~\cite{jensen2016algorithm}); 
ViSQOL~\cite{chinen2020visqol}, an intrusive perceptual metric that estimates mean opinion score based on spectral similarity;
Word Error Rate (WER) comparing the recognition result of Whisper v2~\cite{radford2023robust} on resynthesized speech versus ground truth transcriptions;
Speaker Similarity (SIM), the cosine similarity between the speaker embedding of ground truth and resynthesized speech extracted by ECAPA-TDNN\cite{desplanques2020ecapa,ravanelli2021speechbrain};
Subjective Mean Opinion Score (MOS) assessing audio naturalness and quality on a scale of 1 to 5, with increments of 1.
We randomly selected 40 sentences from the test set and collected 10 ratings for each model for each sample.
We followed the standard approach to collect ratings through AmazonTurk~\cite{naderi2020open,ribeiro2011crowdmos} with the reward of \$0.05 per rating.
Each worker rated 8 sentences and for each sentence audio from 6 different systems were presented.
We reported the average rating for each system on a 95\% confidence interval.

\end{document}